\documentstyle[prl,aps,epsf,twocolumn]{revtex}

\draft

\title
{
Berry's Phase Induced Bose-Einstein Condensation into a Vortex State
}

\author
{
M. Olshanii and M. Naraschewski 
}
\address
{
Jefferson Laboratory, Department of Physics,
Harvard University, Cambridge, MA 02138\\
Institute for Theoretical Atomic and Molecular Physics,
Harvard-Smithsonian Center for Astrophysics, Cambridge, MA 02138
}

\date{\today}

\begin{document}
\maketitle
\begin{abstract}
The existence of a geometric phase in magnetic traps
can be used to Bose condense a magnetically trapped atomic
gas into a vortex state. We 
propose an experimental setup where a magnetic trap together
with a blue detuned laser beam form a multiply
connected trap geometry. The local variation of the 
magnetic quantization axis 
induces a geometric or Berry's phase that allows the atoms to acquire
an effective gauge charge interacting with the analog of a 
magnetic solenoid. It is shown that the ground state 
of such a system may be given by a vortex state. 
We also discuss the influence of atomic interactions on the proposed vortex
production scheme in the context of present
Bose-Einstein condensation experiments with dilute gases.
\end{abstract}


The first experimental realization of Bose-Einstein condensation in
dilute alkali gases 
\cite{BEC_first_experiments_1,BEC_first_experiments_2,BEC_first_experiments_3}
has dramatically revived the interest in weakly interacting Bose gases. 
Compared to previous work, the spatial
inhomogeneity and finite size of the new experimental systems as well as the
possibility to study complicated nonequilibrium dynamics have added many
interesting conceptual elements and have even shed new light
on some basic paradigms of Bose-Einstein condensation. By now, also these novel
Bose condensed systems seemed to be fairly well understood. However,
as we will show in this paper, further surprises may still be abound. 

From the very beginning, the main focus of the investigation of Bose
condensed
alkali gases concentrated on the condensate itself, the macroscopically
occupied ground state of the system. According to a credo of many
textbooks on quantum mechanics (cf.\ \cite{Landau}) the ground state of a
particle in a spherically or axially symmetric potential
has vanishing angular momentum. This statement is usually derived from
the famous theorem of nodelessness of the ground state
\cite{Peierls} that in turn is based on the
assumption that the eigenstates of the Hamiltonian can always be chosen
to be real. However, this assumption
does not necessarily hold for a charged particle in the presence of a
magnetic field \cite{Landau_2}. In our paper we show,
that the action of an effective gauge field on neutral atoms in magnetic
traps, caused by a geometric phase effect, can lead to a situation where
the {\em ground} state of the system
has a nonvanishing angular momentum and the gas
can directly Bose condense into a vortex state.
In so far, our vortex production scheme differs drastically from previous
proposals that suggested the creation of a macroscopically occupied {\em
excited}
state of the system by opto-mechanical stirring \cite{bolda}, adiabatic
population transfer \cite{zoller} or the accidental generation of vortices
in a quench \cite{anglin,drummond}. 

Before discussing the specific details of our proposed trap configuration, we
want to illustrate in terms of classical physics the basic mechanism that
is used to create a ground state with nonvanishing angular momentum.
Consider an electrically charged pendulum in the absence of gravity whose 
pivot point 
is pierced by an infinite solenoid perpendicular to the pendulum's plane of 
oscillation. The Hamiltonian for the pendulum then reads
\begin{eqnarray}
{\cal H} = \frac{M R^2 \dot{\phi}^2}{2}
         = \frac{({\cal L}_{z} - \Delta {\cal L}_z)^2}{2I} \,\, ,
\label{Classical_Hamiltonian}
\end{eqnarray}
where $M$ and $R$ are the mass and length of the pendulum respectively.
$I = M R^2$ is the momentum of inertia of the pendulum, while
${\cal L}_{z} = M R^2 \dot{\phi} + \Delta {\cal L}_z$ is the angular momentum,
canonically conjugated to the phase $\phi$. The  
shift $\Delta {\cal L}_z$ of the angular momentum is given by
\begin{eqnarray}
 \Delta {\cal L}_z = \frac{q\,\Phi_{B}}{2\pi c} \,\, ,
\end{eqnarray}
where $q$ is the pendulum charge, $c$ is the speed of light,
and $\Phi_{B}$ is the magnetic flux across the solenoid.

Apparently, the ground state of the Hamiltonian (\ref{Classical_Hamiltonian}) 
posesses a nonvanishing canonical
angular momentum ${\cal L}_z = \Delta {\cal L}_z$. 
In classical mechanics this increase of the canonical angular momentum
does not lead to any immediately observable physical effect, since 
the angular frequency $\dot{\phi} = 0$ of the pendulum
is the same as in the absence of a magnetic field. Only when the
magnetic field within the solenoid is switched off, the conserved canonical
angular momentum is transformed into mechanical angular momentum and the
pendulum is set into motion. Microscopically this generation of
mechanical angular momentum is a consequence of the displacement current
that is induced by a time-dependent magnetic field due to the Maxwell equation
$\mbox{\boldmath $\nabla$} \times {\bf E} = -(1/c) \dot{\Phi}_{B}$.
In quantum mechanics, however, the canonical 
angular momentum is directly observable without altering the magnetic
field. Here, the 
shift $\Delta {\cal L}_z$ can be made visible by an interference experiment
similar to the one used to display the Aharonov-Bohm effect 
\cite{Aharonov-Bohm}. 

With this brief illustration we have shown how the existence
of a magnetic field can lead to a ground state with nonvanishing angular 
momentum. The direct use of the outlined idea in present experiments with 
Bose gases is however flawed by the electric neutrality of atoms.
Instead, we show that the local variation of the hyperfine spin 
orientation in the inhomogeneous field of a magnetic trap leads to a
geometric phase effect that can be interpreted as the presence
of an effective magnetic field. 

Consider an atom in a 
hyperfine state $F$, which moves in 
an external magnetic field $B({\bf r})$.
The Hamiltonian of the system is given by
\begin{eqnarray}
\label{Hamiltotian}
{\cal H} &=& \frac{\hat{\bf p}^2}{2M} + \hat{V}({\bf r}) \\
\hat{V}({\bf r}) &=& g\,\mu_{B}\,\hat{\bf F} \cdot {\bf B}({\bf r}),
\end{eqnarray}
with the atomic mass $M$, the Bohr magneton $\mu_{B}$, and the 
Lande factor $g$.

If the atomic motion is slow enough, the evolution of the system can 
be described using  
the Born-Oppenheimer approximation. 
The latter assumes that
for every point ${\bf r}$ the internal state 
of the atoms is close to a local eigenstate $|\tilde{m}({\bf r})\rangle$
\begin{eqnarray}
\hat{V}({\bf r}) |\tilde{m}({\bf r})\rangle = \tilde{m}
|\tilde{m}({\bf r})\rangle \, 
\label{tilde_m}
\end{eqnarray}
of the magnetic contribution $\hat{V}({\bf r})$ to the Hamiltonian.
The total wave function can then be written
\begin{eqnarray}
\langle {\bf r}, m | \Psi\rangle =
 \Psi_{\tilde{m}}({\bf r}) \langle m |\tilde{m}({\bf r})\rangle
\label{B-O_wave-function}
\end{eqnarray}
where $|m\rangle$ corresponds to an eigenstate of the projection operator
$F_z={\bf e}_z \cdot \hat{\bf F}$ of the hyperfine spin on the $z$-axis. 

Applying the full Hamiltonian Eq.\ (\ref{Hamiltotian}) to the  
wave function $\langle {\bf r}, m | \Psi\rangle$ of 
Eq.\ (\ref{B-O_wave-function}) shows that the evolution of the 
adiabatic wave function $\Psi_{\tilde{m}}$ is governed by the 
Born-Oppenheimer Hamiltonian \cite{Gauge_Potentials} 
\begin{eqnarray}
H = \frac{[\hat{\bf p} - {\bf A}_{\tilde{m}}({\bf r})]^2}{2M} 
+ V_{\tilde{m}}({\bf r}) + \phi_{\tilde{m}}({\bf r}) \, ,
\label{Hamiltonian_2}
\end{eqnarray}
with the magnetic potential
\begin{eqnarray}
V_{\tilde{m}}({\bf r}) =  g\,\mu_{B}\,\tilde{m}\, |{\bf B}({\bf r})|,
\label{V_potential}
\end{eqnarray}
the vector gauge potential
\begin{eqnarray}
{\bf A}_{\tilde{m}}({\bf r}) = i \hbar \langle \tilde{m} | {\bf
\nabla} \tilde{m} \rangle,
\label{A_potential}
\end{eqnarray}
and the scalar gauge potential
\begin{eqnarray}
\phi_{\tilde{m}}({\bf r}) = 
 \frac
 {
  \hbar^2
  \sum_{\tilde{m}^{\prime} \ne \tilde{m}} 
  |\langle \tilde{m}^{\prime} | {\bf \nabla} \tilde{m} \rangle|^2
 }
 {
 2M
 }.
\label{phi_potential}
\end{eqnarray}
In addition to the phase change caused by the kinetic energy and the potential
energy $V_{\tilde m}({\bf r})$ the adiabatic wave function 
$\Psi_{\tilde{m}}({\bf r})$
acquires the Berry's phase \cite{Berry's_Phase}
\begin{equation}
\label{Berry-phase}
\gamma_{\tilde{m}} 
= \frac{1}{\hbar}\oint_\Gamma\!d{\bf r}\cdot{\bf A}_{\tilde{m}}({\bf r})
\end{equation}
on a semiclassical trajectory along the closed path $\Gamma$.

However, according to Eq.\ (\ref{B-O_wave-function})
there exists an ambiguity in the definition of the adiabatic
wave function $\Psi_{\tilde{m}}({\bf r})$ and of the gauge potentials.
The eigenstate equation (\ref{tilde_m}) defines the 
local eigenstate $|\tilde{m}({\bf r}) \rangle$ only up to
an arbitrary gauge transformation
$
|\tilde{m}({\bf r}) \rangle \longrightarrow 
\exp\lbrack i\varphi_{\tilde{m}}({\bf r}) \rbrack \, |\tilde{m}({\bf r}) 
\rangle
$.
To eliminate this ambiguity we specify the 
gauge by choosing the following definition for  $|\tilde{m}({\bf r}) \rangle$:
\begin{eqnarray}
| \tilde{m}({\bf r}) \rangle  =
 \hat{\cal R}
 \left(
     \frac
       {
         \lbrack {\bf e}_z \times {\bf B}({\bf r}) \rbrack
       }
       {
          |\lbrack {\bf e}_z \times {\bf B}({\bf r}) \rbrack|
       },
     \, \mbox{ang} ({\bf e}_z,{\bf B}({\bf r}))
 \right)
 |m =  \tilde{m} \rangle.
\label{tilde_m_2}
\end{eqnarray}
The operator $\hat{\cal R}({\bf n}, \Theta)$ describes a three-dimensional
rotation around an axis ${\bf n}$ by the angle $\Theta$.
The state $|m = \tilde{m} \rangle$ is an eigenstate of the hyperfine spin
projection on the $z$-axis with eigenvalue $\tilde{m}$.

Our next task is to evaluate the potentials of Eqs.\
(\ref{V_potential}), (\ref{A_potential}), and (\ref{phi_potential}) 
for some particular magnetic trap.
We choose a two-dimensional version of the Ioffe-Pritchard trap
\begin{eqnarray}
{\bf B}({\bf r}) =
   \lambda x \, {\bf e}_x
 - \lambda y \, {\bf e}_y
 + B_{0} \, {\bf e}_z,
\label{B-field}
\end{eqnarray}
assuming that the atomic motion is constrained to 
the $xy$-plane by some auxiliary external field that is independent of the
internal hyperfine spin state.
We will discuss a particular version of such a confinement below.
We also restrict ourselves to the internal hyperfine state
$F = 1$, $\tilde{m} = -1$, $g < 0$ that
corresponds to present experiments with trapped sodium gases. 
The discussion of the general case is postponed
to the end of the paper \cite{Sukumar}.
Using polar coordinates  
for the $xy$-components of $\bf r$, we obtain the expressions
\begin{eqnarray}
V_{-1}(\rho) &=& -g\,\mu_{B}\,\sqrt{B_0^2+\lambda^2\rho^2},
\\
{\bf A}_{-1}(\mbox{\boldmath $\rho$}) &=&
\frac{\hbar \, \Delta {\cal L}_{-1}(\rho) }{\rho}\,
{\bf e}_{\varphi}(\mbox{\boldmath $\rho$}), 
\\
\Delta {\cal L}_{-1}(\rho) &=& -1 + \frac{B_0}{\sqrt{B_0^2 + \lambda^2\rho^2}},
\end{eqnarray}
and
\begin{eqnarray}
\phi_{-1}(\rho) = 
\frac
 {
   2B_0^2\,\lambda^2 + \lambda^4\rho^2
 }
 {
   4(B_0^2 + \lambda^2\rho^2)^2
 },
\end{eqnarray}
with $\mbox{\boldmath $\rho$}=x\,{\bf e}_x+y\,{\bf e}_y$ and
$ {\bf e}_{\varphi}(\mbox{\boldmath $\rho$}) = (-y\,{\bf e}_x + x\, 
{\bf e}_y)/\rho$.
In the special case of a vanishing bias field $B_0 = 0$, 
the angular momentum shift is equal to $\Delta {\cal L}_{-1} = -1$ 
and the gauge vector field 
${\bf A}_{-1}(\mbox{\boldmath $\rho$})$ is equivalent to the
electromagnetic vector field of a solenoid along the $z$-axis with
$q\,\Phi_{B}/c=-2\pi$.
In so far,
the situation of the trapped atoms resembles the previous example of
the electrically charged pendulum interacting with a solenoid. 
We therefore 
conjecture that the ground state of the trapped atom gas has a
nonvanishing canonical angular momentum.
However, before coming to a definitive conclusion about the ground
state of the system, we have to add a few elements to the model
outlined in Eq.\ (\ref{B-field}) in order to make it realistic.

The system Hamiltonian (\ref{Hamiltonian_2})
does not yet contain a longitudinal confinement along
the $z$-direction. We therefore suggest to add a red-detuned light sheet 
parallel to the $xy$-plane. It has to be created by
a far detuned and linearly polarized light beam 
in order to achieve an ac Stark shift that is
independent of the direction of the hyperfine spin ${\bf F}$.
To a good approximation such a dipole trap can be described by a 
harmonic potential with oscillation frequency $\omega_z$. 

Since it is our plan to create a vortex in the absence of a bias field $B_0$, 
we have to find a way to suppress atom losses due to Majorana spin
flips in the center of the trap. Therefore another, far blue-detuned
light beam should be added 
that essentially plugs the region with the most critical trap
losses (cf.\ \cite{BEC_first_experiments_2}).
This linearly polarized light beam is assumed to have a potential energy
$
V_{\rm plug}(\rho) = V_{{\rm plug}, 0} \exp(-2\rho^2/w_{\rm plug}^2)
$
with beam waist $w_{\rm plug}$.

Finally, we have to account for interactions between the gas
atoms. For dilute gases this can be done by adding the mean-field pressure
\begin{eqnarray}
V_{\rm mf}({\bf r}) = gN|\Psi_{-1}({\bf r})|^2
\end{eqnarray}
to the Hamiltonian (\ref{Hamiltonian_2}). The coupling constant
$g = 4 \pi \hbar^2 a / M$ is given by the $s$-wave scattering length
$a$, while $N$ is the total number of atoms. Furthermore we assume
the axial
confinement to be much stronger than the atomic interaction energy 
$(\hbar \omega_z \gg gN|\Psi_{-1}|^2)$.
In this case the atomic wave function  
\begin{equation}
\Psi_{-1}({\bf r}) = \psi_{-1}(\mbox{\boldmath $\rho$}) \cdot \chi^{(0)}(z)
\end{equation}
factorizes
into a radial part $\psi_{-1}$ and the ground state of the axial potential
$\chi^{(0)}(z)$. 
The radial wave function $\psi_{-1}(\mbox{\boldmath $\rho$})$
is then determined by a two-dimensional 
Gross-Pitaevskii Hamiltonian with an effective coupling strength
$g_{\rm 2D} = g\sqrt{M\omega_z}/\sqrt{2\pi\hbar}$.
Note, however, that the existence of such an effectively two-dimensional 
situation is not instrumental for the proposed vortex creation. It was 
solely assumed to facilitate a numerical treatment of the problem.

As a consequence,
the ground state of the system is given as a solution of the
two-dimensional Gross-Pitaevskii equation
\begin{eqnarray}
H_{\rm GP}\, \psi_{-1} = \mu\, \psi_{-1}, 
\label{GPE}
\end{eqnarray}
with the nonlinear Hamilton operator
\begin{eqnarray}
H_{\rm GP} &=&   
   \frac{[\hat{\bf p} - {\bf A}_{-1}(\mbox{\boldmath $\rho$})]^2}{2M}
   + V_{-1}(\rho)
   + \phi_{-1}(\rho)
\nonumber\\
&&
   + V_{\rm plug}(\rho)
   + g_{\rm 2D} N |\psi_{-1}(\mbox{\boldmath $\rho$})|^2 + \frac{\hbar\omega_z}
{2},
\end{eqnarray}
the chemical potential $\mu$, the total atom number $N$.
Our goal is to prove numerically that the minimum of the system energy
\begin{eqnarray}
E\lbrack\psi,\psi^{\star}\rbrack =
\langle \psi
|
  N H_{\rm GP}
  - \frac{g_{\rm 2D}\,N^2 |\psi_{-1}|^2}{2}
|\psi
\rangle .
\end{eqnarray}
is reached for a state with  
nontrivial value of the angular momentum ${\cal L}_z = -1$.
Using the separability of the radial and angular motion, 
we divide the energy minimization procedure into two steps:
first, we minimize the energy under the constraint of a
particular value of the angular momentum ${\cal L}_z$, then
we compare the obtained energies in order to find its global minimum.

\begin{figure}
\begin{center}
\leavevmode
\epsfxsize=0.45\textwidth
\epsffile{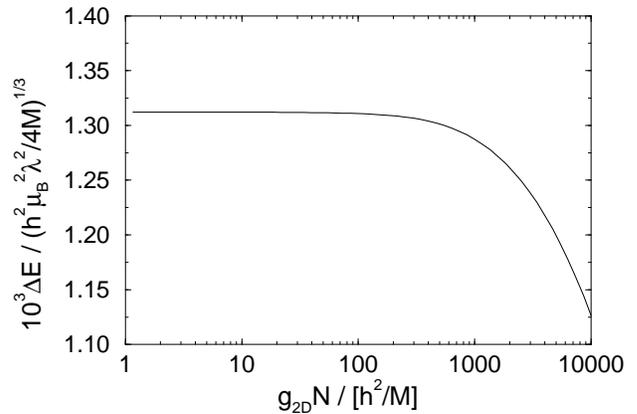}
\end{center}
\caption
{
\label{Efig}
In this figure, we plot for sodium atoms the energy difference $\Delta E =
E_{{\cal L}_z = 0} - E_{{\cal L}_z = -1}$ between energy minimizing states with
constrained angular momenta ${\cal L}_z=0$ and ${\cal L}_z=-1$
as a function of the atomic interaction strength $g_{\rm 2D} N$.
We use the two-dimensional Ioffe-Pritchard trap with optical plug and
confining light sheet that is described in the body of this paper. 
We use the realistic set of trap parameters
$\lambda = 223 \, {\rm Gs}/{\rm cm}$, 
$w = 3.05 \, \mu{\rm m} \, (= 10 \times (2\hbar^2/\mu_B \lambda M)^{1/3})$,  
$V_{{\rm plug}, 0} = 2\pi \times \hbar \times 9.7 {\rm MHz}$,
and $\omega_{z} = 26 \, {\rm kHz}$. 
For this set of parameters, 
the interaction strength $g_{2D}NM/\hbar^2 = 1000$ 
corresponds to $N = 1250$ atoms and 
an axial Thomas-Fermi parameter of
$g_{\rm 2D}\bar{n}_{\rm 2D}/\hbar \omega_{z} \approx 0.5$, where 
$\bar{n}_{\rm 2D}$ is the peak value of the two-dimensional density 
in the $xy$-plane.
}
\end{figure}

\begin{figure}
\begin{center} 
\parbox{4cm}
{  
\begin{center}
\leavevmode
\epsfysize=0.16\textwidth
\epsffile{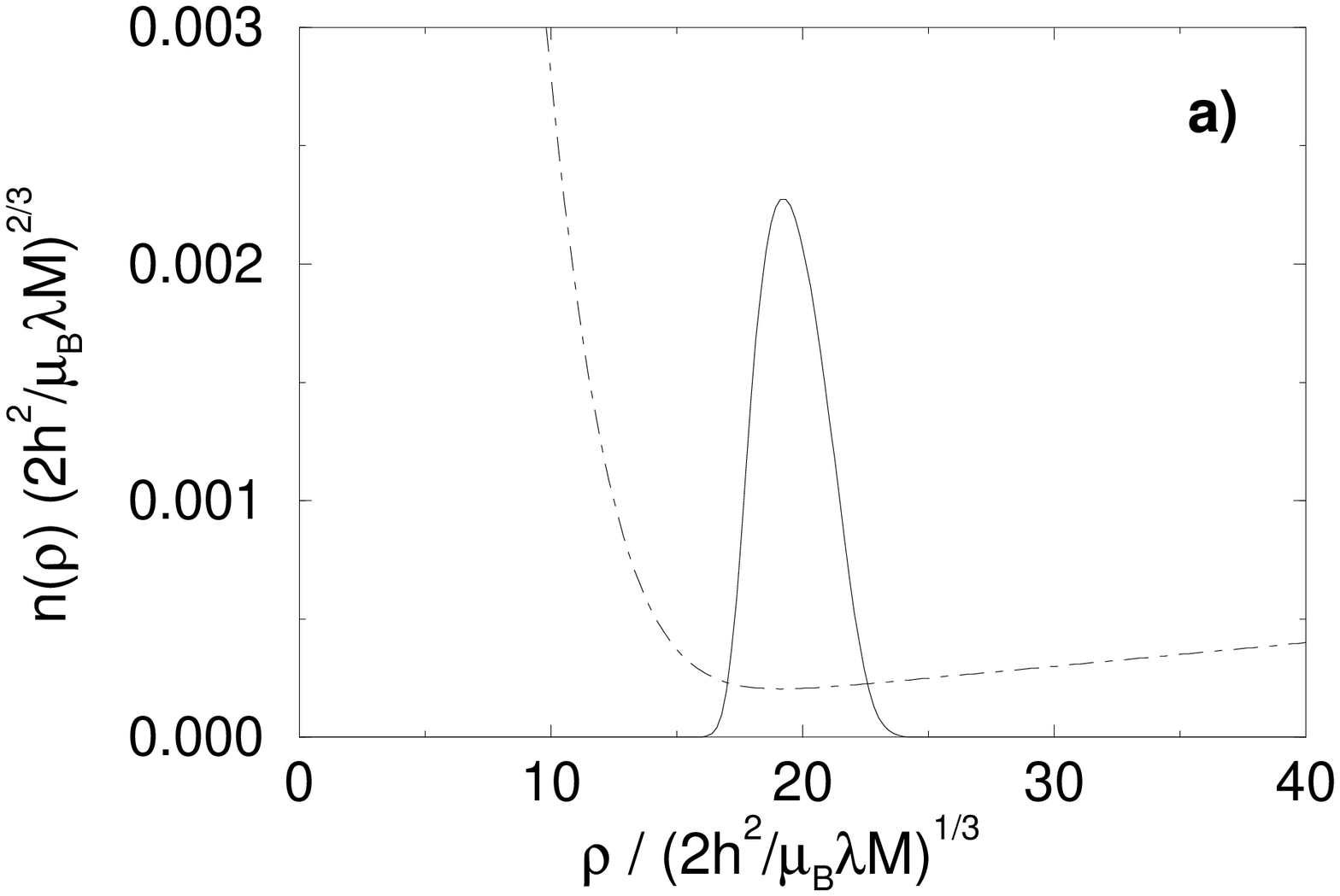}
\end{center}
}  
\parbox{4cm}
{  
\vspace*{1mm}
\begin{center}
\leavevmode
\epsfysize=0.152\textwidth
\epsffile{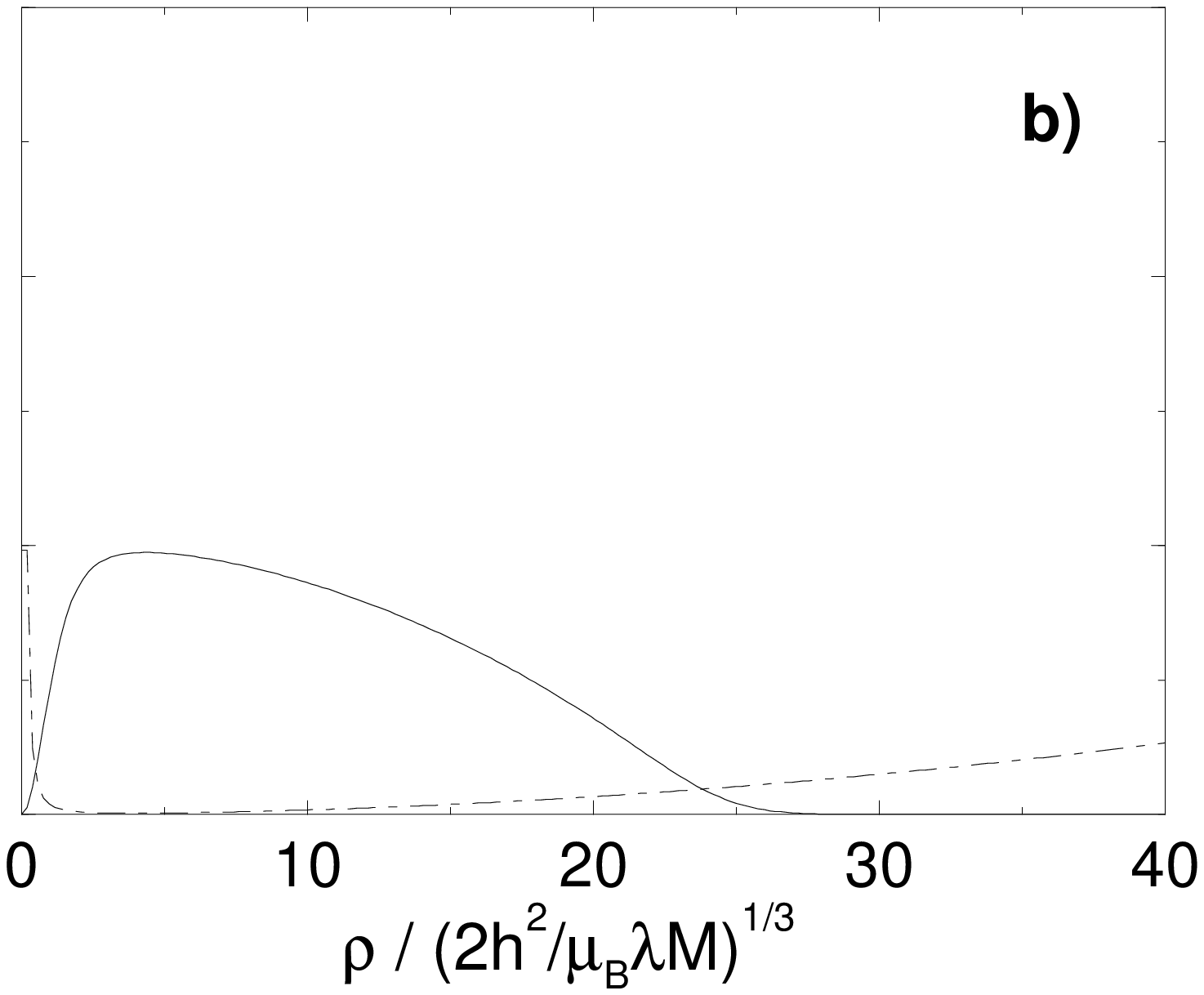}
\end{center}
}  
\end{center}    
\caption
{
\label{densityfig}
Here, we show the normalized two-dimensional density distribution
$n(\rho)$ for a vortex state with ${\cal L}_z=-1$ and 1250 sodium atoms
($g_{2D}NM/\hbar^2 = 1000$). 
The left plot corresponds to the interacting ground state of the 
toroidal trap described in the caption of Fig.\ \ref{Efig}.
The dot-dashed line shows the effective potential energy including 
magnetic field, light plug, gauge potentials, and the cenrifugal barrier.
After the Bose-Einstein condensation into the vortex state is completed,  
a bias field $B_0 = 2 \, {\rm Gs}$ is slowly turned on and 
the optical plug is slowly removed. 
The resulting trap configuration consists of a two-dimensional 
Ioffe-Pritchard trap with radial frequency
$\omega_{\perp} = 391 \, {\rm Hz}$, which corresponds to 
$0.08$ in the units of Fig.\ \ref{Efig},
and the axial light sheet described in the caption of Fig.\ \ref{Efig}. 
The right figure demonstrates that the resulting atomic density distribution
vanishes for small $\rho$ as a consequence of the nonvanishing centrifugal
barrier of a vortex state.
}
\end{figure}

As it was expected, we found that for the case of a vanishing
bias field $B_0 = 0$, the energy minimum is reached for 
the ${\cal L} = -1$ state. As an illustration, we plot the energy difference 
between the ${\cal L} = -1$ and ${\cal L} = 0$ state in
Fig.\ \ref{Efig}. One can see that 
for a broad range of condensate populations the energy of the vortex state is 
constantly lower than the energy of the axially symmetric state. 
As a further illustration, Fig.\ \ref{densityfig}a shows the radial 
distribution of the atomic density of the vortex state. 
We regard this as a convincing proof of a possible Bose-Einstein condensation 
into a vortex state.
It is no surprise, however, that the described effect has not yet been 
observed experimentally, since
for typically used large bias fields of $B_0 \gg \lambda\bar{\rho}$,
$\bar{\rho}$ being a typical cloud radus, the condensation into a vortex state
is energetically not favored.

The main
inconvenience of the proposed setup lies in the potential difficulties
that it poses for the detection of a vortex state. The density distributions
for different angular momenta are almost indistinct, since 
the most obvious signature of vorticity -- a hole in the center of the 
density distribution -- is masked by the optical plug. 
We therefore suggest the following experimental sequence in order to make the
vorticity directly observable.
\begin{itemize}
\item 
Turn on slowly a bias field $B_0$ that is strong enough to prevent trap losses
by Majorana spin flips.
This stage is equivalent to turning off the solenoid field in the case of the 
pendulum \cite{dm/dt_terms}.
\item
Remove the optical plug adiabatically, by lowering the light intensity
(cf.\ Fig.\ \ref{densityfig}b).
Then, remove the magnetic confinement instantaneously
and allow the cloud to expand freely within the $xy$-plane.
\end{itemize}
A detectable hole in the middle of the density distribution can now be taken 
as a distinctive feature of a vortex state.
Alternatively, the vorticity of the state can be determined by 
an interference experiment \cite{bolda2}.

In conlusion, we have pointed out that magnetically trapped atoms are
subject to an effective gauge potential that is related to the existence of
a geometric phase. Under certain circumstaces, 
the trapped atoms behave as if they were charged
particles interacting with a magetic field of a  flux carrying solenoid. 
We show that this
effective
magnetic field can yield a ground state of the system with nonvanishing 
angular momentum, and thereby open the possibility of a direct
Bose-Einstein condensation into a vortex state.
Our results can be easily generalized to an arbitrary
value of the hyperfine spin $F$ and to different kinds of magnetic
traps. Indeed, one can show that the angular momentum of the
ground state is then given by
\begin{eqnarray}
{\cal L}_z = -\tilde{m}\Upsilon,
\end{eqnarray}
where $\Upsilon$ is the topological index of the used magnetic field
configuration. It measures in terms of $2\pi$ the 
rotation
angle of the magnetic field vector ${\bf B}({\bf r})$ along a closed path
around the center of the trap. The proposed trap geometry of this
paper corresponds to a topological index
of $\Upsilon = -1$. In contrast, a quadrupole 
trap combined with an additional red-detuned light sheet in the $xy$-plane
would correspond to an index $\Upsilon = +1$.


After this work has been materially completed, we learned about a related 
vortex production scheme employing the Ahoronov-Casher effect \cite{you}.
It combines a trap geometry very similar to ours with an electrically charged
wire. In principle, such a scheme would 
allow the controlled creation of arbitrary vortex states, even
though
the experimental requirements seem to be challenging. We show, in 
contrast, that Bose-Condensation into a vortex state may already occur
in a more basic trap setup without the charged wire.

M.O.  was supported by the National Science Foundation
grant for light force dynamics \#PHY-93-12572. 
M.N. acknowledges support by the Deutsche Forschungsgemeinschaft.
This work was also partially funded by the NSF through
a grant for the Institute for Theoretical Atomic and Molecular
Physics at Harvard University and the Smithsonian Astrophysical
Observatory.




%
%

%

\begin{thebibliography}{99}
\bibitem{dm/dt_terms}
Generally speaking for the case of the time-dependent magnetic field
we should add another Berry's phase-like term
$-i\hbar \langle \tilde{m} | (\partial/\partial t) \tilde{m} \rangle$
to the Hamiltonian (\ref{GPE}). In our particular case this term vanishes.
\end{thebibliography}
\end{document}